\documentclass{aa}
\usepackage{times}

\newcommand{\ltsima} {$\; \buildrel < \over \sim \;$}
\newcommand{\simlt}  {\lower.5ex\hbox{\ltsima}}            
\newcommand{\gtsima} {$\; \buildrel > \over \sim \;$}
\newcommand{\simgt}  {\lower.5ex\hbox{\gtsima}}            
\newcommand{\ferg}{erg cm$^{-2}$ s$^{-1}$ }

\newcommand{\be} {\begin{equation}}

\newcommand{\ee} {\end{equation}}

\newcommand{\RXTE}{{\em Rossi}XTE}
\newcommand{\xmm}{{\em XMM-Newton}}
\newcommand{\bc}{\begin{center}}
\newcommand{\ec}{\end{center}}

\def \hcm {\hbox {\ifmmode $ atoms cm$^{-2}\else atoms cm$^{-2}$\fi}}

\def\ee {1E~1048.1--5937~}

\begin{document}



\title
{Three \xmm\ observations of the Anomalous X--ray Pulsar \ee :
long term variations in spectrum and pulsed fraction\thanks{Based on 
observations obtained with \xmm, an ESA science mission with instruments 
and contributions directly funded by ESA Member States and NASA}
}

\author{ A.~Tiengo\inst{1,2}, S.~Mereghetti\inst{1}, R.~Turolla\inst{3},
S.~Zane\inst{4}, N.~Rea\inst{5,6}, L.~Stella\inst{5},
G.L.~Israel\inst{5}}

\institute {
{1) Istituto di Astrofisica Spaziale e Fisica Cosmica, Sezione di Milano ``G.Occhialini'',
via Bassini 15, I-20133 Milano, Italy}
\\
{2) Universit\`a degli Studi di Milano, Dipartimento di Fisica,
via Celoria 16, I-20133 Milano, Italy}
\\
{3) Universit\'a di Padova, Dipartimento di Fisica, via Marzolo 8,
I-35131 Padova, Italy}
 \\
{4) Mullard Space Science Laboratory, University College London,
Holmbury St. Mary, Dorking Surrey, RH5 6NT, UK}
\\
 {5) INAF, Osservatorio Astronomico di Roma, via
Frascati 33, I-00040 Monteporzio Catone, Roma, Italy}
\\
 {6) SRON - National Institute for Space Research, Sorbonnelaan, 2, 3584 CA, 
Utrecht, The Netherlands}
 }

\offprints{A. Tiengo, tiengo@mi.iasf.cnr.it}

\date{Received / Accepted}

\authorrunning{A. Tiengo et al. }
\titlerunning{ \ee}

\abstract{We report the results of a recent (July 2004)  \xmm\
Target of Opportunity observation  of the Anomalous X--ray pulsar
\ee\, together with a detailed re-analysis of previous
observations carried out in 2000 and 2003. In July 2004 the source
had a 2-10 keV flux of 6.2$\times$10$^{-12}$ erg cm$^{-2}$
s$^{-1}$ and a pulsed fraction P$_F$=0.68. The comparison of the
three data sets shows the presence of an anti-correlation between
flux and pulsed fraction, implying that previous estimates of the
source energetics based on the assumption of a large and constant
pulsed fraction might be significantly underestimated. The source
spectrum is well described by a power law plus blackbody model
(kT$\sim$0.63 keV, photon index $\Gamma$$\sim$2.7--3.5) or,
alternatively, by the sum of two blackbodies of which the hotter
is Comptonized by relativistic electrons. In this case the
temperatures are kT${_1}\sim$0.2--0.3 keV and kT${_2}\sim$0.4--0.5
keV and the emitting area of the cooler component is consistent
with the whole neutron star surface.  The long term luminosity
variation of a factor $\ga$2 is accompanied by relatively small
variations in the spectral shape. Phase resolved spectroscopy
indicates a harder spectrum in correspondence of the pulse
maximum. No spectral features have been detected with 4$\sigma$
limits on the equivalent width in the range $\sim$10-220 eV,
depending on line energy and width.
 \keywords{Stars: individual: \ee -- X-rays: stars  } }

\maketitle

\section{Introduction}

The Anomalous X-ray Pulsars (AXPs, Mereghetti \& Stella 1995; van
Paradijs, Taam \& van den Heuvel 1995) were originally considered
a subclass of the accreting X-ray pulsars characterized by spin
period of a few seconds, stable spin-down,  soft X-ray spectrum
and luminosity larger than the rotational energy loss. The absence
of bright optical counterparts and of Doppler modulations of their
pulse period has then led to the conclusion that they are very
likely isolated neutron stars (see Mereghetti et al. 2002 for a
review). Although the possibility that they are powered by
accretion  from a residual disk cannot be completely excluded,
the model currently considered most successful involves highly
magnetized neutron stars (``magnetars''), in which the decay of an
intense magnetic field (10$^{14}$--10$^{15}$ G) powers the X-ray
emission. This model was first proposed (Duncan \& Thompson 1992)
to explain a different class of enigmatic objects, the Soft
Gamma-ray Repeaters (SGRs) and then extended to the AXPs due to
their many common properties (Thompson \& Duncan 1996), including
the recent observations of bursting activity in a couple of AXPs
(Gavriil, Kaspi \& Woods 2002; Kaspi et al. 2003).

\ee is a key object to understand the connection between AXPs and
SGRs and to study the physical processes involved in the X-ray
emission from these sources. Besides being the first AXP from
which X-ray bursts were discovered, other properties make 1E
1048.1-5937 a possible transition object between AXPs and SGRs: it
is one of the AXPs with the hardest X-ray spectrum and with the most
variable period evolution, which are typical characteristics of
the two best studied SGRs (SGR 1900+14 and SGR 1806--20).

Flux variations were reported for \ee in the past, comparing
measurements from different satellites, which could be affected by
large systematic uncertainties (Oostebroek et al. 1998). Only
recently, the fact that \ee is a variable X-ray source has been
firmly established, when very different source intensities were
registered in two {\it XMM-Newton} and two {\it Chandra}
observations (Mereghetti et al. 2004). This variability  has been
confirmed by the monitoring program carried out with {\it
Rossi-XTE} which showed two extended flares in its pulsed
flux\footnote{due to the difficulty of background subtraction and
the presence of the bright and variable X-ray source Eta Carinae
in the field of view of this non-imaging satellite, the RXTE data
cannot give accurate information on the total (pulsed plus
unpulsed) flux level.} (Gavriil \& Kaspi 2004). Since the distance of
\ee is not known (see e.g. Gaensler et al. 2005), we assume a 
distance of 5 kpc. Following
this assumption, the 2--10 keV luminosity of \ee has been observed to vary
between $\sim$10$^{34}$ and $\sim$10$^{35}$ erg s$^{-1}$.

The \xmm\ satellite has observed \ee in three occasions: a short
snapshot observation was done in December 2000 (Tiengo et al.
2002); a 10 times longer observation was performed in June 2003,
when the source was more than a factor 2 brighter (Mereghetti et
al. 2004);  finally, a Target of Opportunity observation was
requested in response to the detection of a burst from the
direction of \ee on 2004 June 29 (Kaspi et al. 2004). These
observations, spanning an interval of more than three years and
catching the source at different intensity states,  offer the
possibility of a systematic study of the long term changes in the
source properties based on a homogeneous set of data. Furthermore
the high statistics obtained in the June 2003 observation allows
the most detailed phase resolved spectral analysis and search for
spectral lines ever carried out for this source.

\begin{table*}[htbp]
\begin{center}
  \caption{XMM-Newton observations of \ee }

    \begin{tabular}[c]{ccccccc}
\hline
Observation & Date   & Duration   &  PN Mode$^{(a)}$ &  MOS1 Mode$^{(a)}$ & MOS2 Mode$^{(a)}$   & Period\\
           &         &            & Net exposure    &  Net exposure    &  Net exposure & Pulsed fraction$^{(b)}$ \\
\hline
A & 2000 Dec 28 &      8 ks     &  FF               & SW               &   SW       &  6.45253$\pm$0.00009 s\\
  &             &                 &  4.5 ks         & 7 ks              &   7 ks    &  (91$\pm$3)\%   \\
\hline
B & 2003 June 16 &     50 ks     &   FF             &   SW               & SW      &  6.454894$\pm$0.000002 s \\
  &             &                 &   40 ks          &  46 ks         &      46 ks  &       (54.6$\pm$0.6)\% \\
\hline
C & 2004 July 08 &    30 ks      &   Ti        &  SW            & Ti    & 6.456109$\pm$0.000005 s\\
  &             &               &     16 ks          &  15 ks         &  15 ks  &  (68$\pm$1)\% \\
\hline
\end{tabular}
\end{center}
\begin{small}
$^{(a)}$ FF = Full Frame (time resolution 73 ms); SW = Small Window
(time res. 0.3 s); Ti = Timing (time res. 0.03 ms (PN), 1.5 ms (MOS))\\
$^{(b)}$ in the 0.6-10 keV energy range. \\
\end{small}
\label{obslog}
\end{table*}

\begin{table*}[htbp!]
\begin{center}
  \caption{Results of phase averaged spectroscopy$^{(a)}$  }
    \begin{tabular}[c]{ccccc}
\hline
Model & Parameter & A & B$^{(b)}$ & C \\
\hline \hline
PL + BB & N$_H$ (10$^{22}$ cm$^{-2}$) & 0.95$\pm$0.09 & 1.08$\pm$0.02 & 1.10$^{+0.06}_{-0.03}$ \\
& kT$_{BB}$ (keV) & 0.63$\pm$0.04 & 0.627$\pm$0.007 & 0.623$^{+0.005}_{-0.006}$\\
& R$_{BB}$ (km)$^{(c)}$ & 0.8$\pm$0.1 & 1.29$\pm$0.03 & 1.04$^{+0.04}_{-0.05}$ \\
& $\Gamma$ & 2.9$\pm$0.2  & 3.27$\pm$0.05 & 3.44$^{+0.09}_{-0.06}$ \\
& PL norm$^{(d)}$ & 3.8$\pm$0.3 & 12.7$^{+0.1}_{-0.3}$ & 9.4$\pm$0.3\\
& $\chi^2_{red}$ (d.o.f.) & 0.963 (255) & 1.046 (519) & 1.041 (432) \\
\hline
BB1 + BB2  & N$_H$ (10$^{22}$ cm$^{-2}$) & 0.55$^{+0.06}_{-0.05}$ & 0.62$\pm$0.01 & 0.67$\pm$0.02 \\
           & kT$_{BB1}$ (keV)& 0.47$^{+0.05}_{-0.06}$ & 0.44$\pm$0.01 &  0.37$\pm$0.02 \\
           & R$_{BB1}$ (km)$^{(c)}$ & 1.7$^{+0.4}_{-0.2}$ & 2.8$\pm$0.1 & 3.1$^{+0.4}_{-0.3}$ \\
           &  kT$_{BB2}$ (keV) & 1.0$^{+0.2}_{-0.1}$ & 0.86$\pm$0.02 & 0.76$^{+0.03}_{-0.02}$ \\
           & R$_{BB2}$ (km)$^{(c)}$ &  0.3$\pm$0.1 & 0.63$\pm$0.05 & 0.7$\pm$0.1 \\
           & $\chi^2_{red}$ (d.o.f.) & 1.004 (255) & 1.365 (519) & 1.118 (432) \\
\hline
CBB & N$_H$ (10$^{22}$ cm$^{-2}$) & 0.53$\pm$0.04 & 0.588$\pm$0.008 & 0.57$\pm$0.02 \\
& kT$_{BB}$ (keV) & 0.40$\pm$0.02 & 0.412$\pm$0.004 & 0.40$\pm$0.01 \\
& R$_{BB}$ (km)$^{(c)}$ & 1.7$\pm$0.1 & 2.75$\pm$0.04 & 2.3$\pm$0.1 \\
& $\alpha$$^{(e)}$ & 3.8$\pm$0.2 & 4.40$\pm$0.06 & 4.4$\pm$0.2 \\
& $\chi^2_{red}$ (d.o.f.) & 0.994 (256) & 1.401 (520) & 1.273 (433) \\
\hline
BB+CBB & N$_H$ (10$^{22}$ cm$^{-2}$) & 0.8$\pm$0.2 & 0.82$\pm$0.02 & 0.79$^{+0.10}_{-0.05}$ \\
& kT$_{BB}$ (keV) & 0.22$^{+0.12}_{-0.05}$ & 0.23$\pm$0.02 & 0.26$^{+0.03}_{-0.04}$ \\
& R$_{BB}$ (km)$^{(c)}$ & 5.5$^{+6.5}_{-3.5}$ & 8$\pm$2 & 5.9$^{+4.2}_{-1.4}$ \\
& kT$_{CBB}$ (keV) & 0.44$^{+0.09}_{-0.05}$ & 0.45$^{+0.02}_{-0.01}$ & 0.48$^{+0.03}_{-0.04}$ \\
& R$_{CBB}$ (km)$^{(c)}$ & 1.5$\pm$0.4 & 2.3$\pm$0.2 & 1.6$^{+0.3}_{-0.1}$ \\
& $\alpha$$^{(e)}$ & 4.1$^{+0.7}_{-0.4}$ & 4.9$\pm$0.1 & 5.4$\pm$0.5 \\
& $\chi^2_{red}$ (d.o.f.) & 0.964 (254) & 1.026 (518) & 1.042 (431) \\
\hline \hline

\end{tabular}
\end{center}
\begin{small}
$^{(a)}$ Errors are at the  90\% c.l. for a single interesting parameter\\
$^{(b)}$ A 2\% systematic error has been applied to the model\\
$^{(c)}$ Radius at infinity assuming a distance of 5 kpc\\
$^{(d)}$ Normalization of the power law component in units of 10$^{-3}$ photons cm$^{-2}$ s$^{-1}$ keV$^{-1}$ at 1 keV\\
$^{(e)}$ Comptonization parameter as defined in the text\\
\end{small}
\label{spectra}
\end{table*}

\begin{figure*}[htb!]
\mbox{}
 \vspace{9.5cm}
 \includegraphics{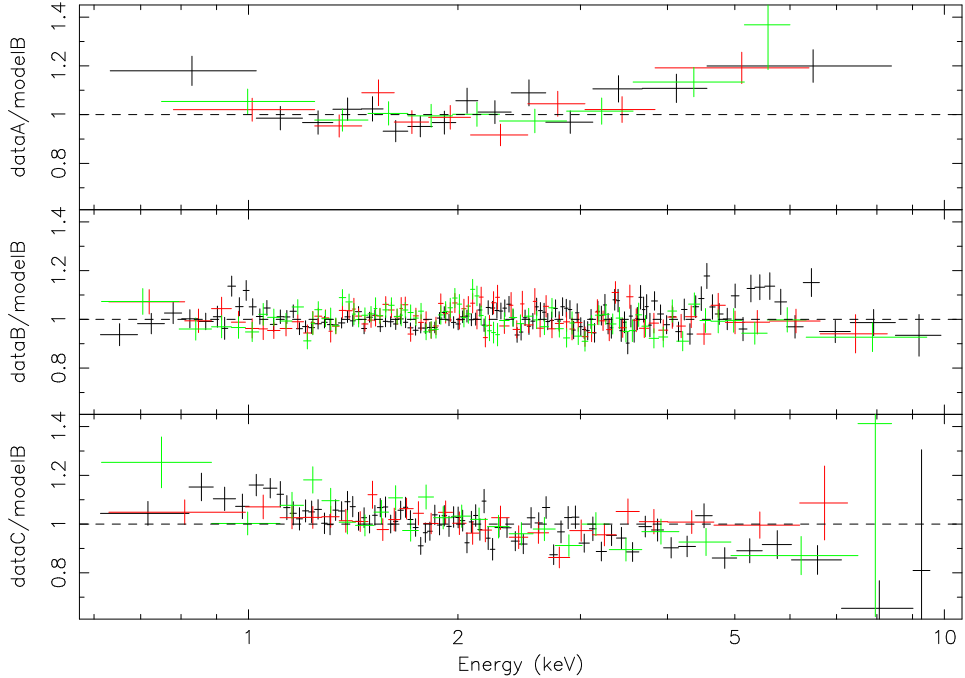}
 \caption[]{Ratios between the spectra of the three observations (A: top, B: middle,
 C: bottom) and the (renormalized) best fit model of observation B.
The colors indicate the three EPIC cameras: PN in black, MOS1 in
red, MOS2 in green. The data have been graphically rebinned to
show more clearly the spectral differences.}
 \label{res}
\end{figure*}

\section{Observations and data analysis}

Table \ref{obslog} gives a log of the three \xmm\ observations of
\ee (hereafter observations A, B and C). The results of obs. C are
reported here for the first time, together with a more detailed
analysis of the previous two observations. We concentrate mostly
on data from the EPIC PN (Str{\"u}der et al. 2001) and MOS (Turner
et al. 2001) cameras since only in obs.B the  Reflection Gratings
Spectrometer (RGS, den Herder et al. 2001) provided a sufficient
number of source photons. In all the observations the medium
thickness optical blocking filter was used for the three EPIC
cameras, while different observing modes were used.  In
particular, during obs. C the PN and MOS2 cameras were operated in
Timing mode, which has a better time resolution (0.03 ms for PN
and 1.5 ms for MOS), but only mono-dimensional imaging capability,
which results in a higher background in the source extraction
region.

All the data have been processed with the same version of the
analysis software (SAS 6.0.0) and using the most recent
calibration files.
Time intervals with high particle background were filtered out.
This reduces the deadtime-corrected exposure time to 16 ks for PN
and 15 ks for MOS in obs. C and 40 ks for PN and 46 ks for MOS in
obs. B. Due to the moderate amplitude of particle background
fluctuations and the short exposure time, no time filter was
applied to the   data of obs. A.

For all the data taken in imaging modes (i.e., PN and MOS data of
obs. A and B and MOS1 data of obs.C), a circle of 30$''$ radius
centered at the position of \ee was used for the source extraction
region for spectral and timing analysis. For the Timing mode data
the source events were selected from a 40$''$ wide strip around
the source position. The regions for the background subtraction
were chosen from the same chip where the source was detected.

No bursts from the source were found by a visual inspection of the
source and background light curves of the three observations.
This is not surprising because, although EPIC has the 
sensitivity to detect bursts like the three observed by \RXTE~ up to now, 
their recurrence time is longer
than the total exposure time of our observations.

\subsection{Spectral analysis}

For each observation,  the 0.6--10 keV spectra from the three EPIC
cameras were rebinned in order to have at least 30 counts per
channel and fitted simultaneously using the  XSPEC package
v11.3.0.
To account for the cross-calibration uncertainties between
different detectors  and operating modes, the relative
normalization was let free to vary.
Fixing  the PN normalization factor to 1, the values for the MOS1
and MOS2 were 1.02$\pm$0.03 and 0.97$\pm$0.03 for obs. A,
1.008$\pm$0.008 and 1.043$\pm$0.009 for obs. B and 1.13$\pm$0.02
and 0.89$\pm$0.02 for obs. C. The larger difference of the
last observation values reflects the greater uncertainties in the
flux calibration of the timing mode used in the MOS2 and PN
detectors.

The statistical quality of the obs. B spectrum is so high
($>$130,000 counts in the PN only) that the deviations from the
best-fit models are dominated by the systematic uncertainties in
the instrumental calibration. Therefore we added in all the model
fits to this data set a 2\% systematic error, perfectly consistent
with the current uncertainty in the EPIC calibration (Kirsch et
al. 2004). In this way formally acceptable   $\chi^2$ values are
obtained for the good fits.

Consistently with previous results on this AXP we found that
simple models based on a single spectral component are not
adequate. Therefore we tried different more complex models,
obtaining the best fit spectral parameters summarized in Table 2.
The ``canonical'' AXP model consisting of an absorbed power law
plus blackbody provides a good fit to all the spectra, with
absorption N$_H$$\sim10^{22}$ cm$^{-2}$, blackbody temperature
kT$_{BB}$$\sim$0.6 keV and photon index $\Gamma$ in the range
$\sim$2.7-3.5.  The source luminosity was different in the three
observations,
with 2-10 keV observed fluxes of 4.7$\times10^{-12}$ erg cm$^{-2}$
s$^{-1}$, 10.0$\times10^{-12}$ erg cm$^{-2}$ s$^{-1}$, and
6.2$\times10^{-12}$ erg cm$^{-2}$ s$^{-1}$, respectively in obs.
A, B, and C.

The sum of two blackbody components, successfully used in the past
for other AXP spectra (e.g., Israel et al. 2001; Halpern \&
Gotthelf 2004), yields acceptable results for obs. A, but is
clearly rejected by the higher quality spectra of the two longer
observations.

In an attempt to physically link the power law and blackbody like
components, we also tried a simple Comptonization
model\footnote{the applicability of this model  to the magnetar
case is discussed in Sect.~3}. The basic idea is that soft thermal
photons, possibly produced at the star surface, may be upscattered
by relativistic $\mathrm{e}^-$ (or $\mathrm{e}^\pm$) of small
optical depth and mean Lorentz factor $\langle\gamma\rangle$. An
analytical, approximated expression which relates the incident to
the emergent intensity is given in Rybicki \& Lightman (1979; see
also references therein) for monochromatic incident radiation of
energy $\mathrm{E}'$, $\mathrm{I_e(E)}\sim
\mathrm{I_i(E')(E/E')}^{1-\alpha}$. In the previous expression
$\alpha=1-\ln\tau^B_{\mathrm{es}}/\ln\mathrm{A}$, where
$\tau^B_{\mathrm{es}}$ is the scattering depth in a magnetized medium
and $\mathrm{A}\sim
4\langle\gamma^2\rangle/3$ is the mean energy amplification factor
per scattering. The case of a blackbody photon input is easily
obtained by convolving the monochromatic expression with the
initial photon distribution.
The photon spectrum is 
given by $\mathrm{CE}^{-\alpha}\int_0^{\mathrm E} \mathrm{dE'\, 
E'}^{1+\alpha}/[\exp{(\mathrm{E'/kT_{BB}}})-1]$.
The model parameters are then the blackbody temperature, normalization 
$C$ and
the photon index $\alpha$.
We find that the fit with this model alone (CBB in Table 2)
is not satisfactory, but the addition of a second, softer
blackbody component, with kT$\sim$0.2-0.3 keV, results in
acceptable fits for all the observations. The hotter Comptonized
blackbody has a smaller temperature and larger emitting area than
that found in the blackbody plus power law model.

Independent of the adopted spectral model, it is clear from the
best fit parameters reported in Table 2 that the
spectrum changed between the different observations. This can also
be seen from Fig.\ref{res}, which shows the ratios between the
spectra of the three observations and the best fit model of obs. B
scaled only in overall normalization (the normalization
factors are reported in the first line of Table 3). 
The spectrum of obs. C is significantly softer 
than the template spectrum, while that of obs. A is slightly harder,
although, due to its poorer statistiscs,
the fit is formally acceptable, as reported in Mereghetti et al. (2004).

For the sake of simplicity we tried to interpret the long term
spectral variations using the model with less parameters, i.e the
blackbody plus power law. In order to check whether the long term
spectral variation could be ascribed to a change in only one or
two of the four\footnote{the overall spectral shape, beside
depending on N$_H$, $\Gamma$ and kT$_{BB}$, is a function of the
relative normalization of the two components} relevant parameters
we fitted the spectra of obs. A and C keeping some of them fixed
to the best fit values of obs. B.
The results are summarized in Table 3, where one can
see that the  variations are described in an almost equivalent
way by a change either in the blackbody temperature (which is
slightly preferred for obs. A) or in the photon index (which gives
a better fit to obs. C). If the photon index and temperature are
kept fixed and only their relative normalization or the N$_H$ free
to vary, a significant improvement in the $\chi^2$ value is
obtained for obs. C but not for obs. A.

The possibility that one of the two spectral components remained constant
(both in shape and normalization) throughout the three observations was
also explored, with negative results. In fact, due to the flux variation 
between the two observations, fitting the spectrum of obs. A with the
parameters of either the power law or the blackbody component fixed to those
of obs.B gives $\chi^2_{red}>$2 (even if the N$_H$ is left free to vary).


\begin{table}[htbp!]
\begin{center}
  \caption{Fits to obs. A and C with a power-law plus blackbody model and
  different sets of fixed parameters.
}
    \begin{tabular}[c]{ccc}
\hline
Parameter$^{(a)}$ & A & C \\
\hline \hline
Normalization factor & 0.432$\pm$0.006 & 0.658$\pm$0.004 \\
 $\chi^2_{red}$ (d.o.f.) & 1.081 (261) & 1.341 (438) \\
\hline
N$_H$ (10$^{22}$ cm$^{-2}$) & 1.07$\pm$0.02 & 1.02$\pm$0.01\\
Normalization factor & 0.429$^{+0.006}_{-0.009}$ & 0.632$^{+0.006}_{-0.004}$ \\
$\chi^2_{red}$ (d.o.f.) & 1.084 (260) &  1.112 (437) \\
\hline
R$_{BB}$ (km) & 0.84$\pm$0.03 & 0.96$\pm$0.02 \\
PL norm & 5.6$\pm$0.2 & 9.3$\pm$0.2 \\
$\chi^2_{red}$ (d.o.f.) & 1.084 (260) &  1.126 (437) \\
\hline
kT$_{BB}$ (keV) & 0.69$\pm$0.02 & 0.602$\pm$0.008\\
R$_{BB}$ (km) & 0.66$^{+0.02}_{-0.06}$ & 1.07$^{+0.04}_{-0.02}$ \\
PL norm & 6.0$\pm$0.2 & 9.0$\pm$0.2 \\
$\chi^2_{red}$ (d.o.f.) & 0.978 (259) &  1.073 (436) \\
\hline
$\Gamma$ & 3.02$^{+0.07}_{-0.06}$ & 3.44$^{+0.04}_{-0.06}$ \\
R$_{BB}$ (km) & 0.72$\pm$0.05 & 1.03$^{+0.02}_{-0.03}$ \\
PL norm & 5.6$^{+0.2}_{-0.3}$ & 9.3$^{+0.2}_{-0.1}$\\
$\chi^2_{red}$ (d.o.f.) & 0.999 (259) &  1.040 (436) \\
\hline \hline

\end{tabular}
\end{center}
\begin{small}
$^{(a)}$ The parameters not indicated in this Table have been
fixed to the best-fit  values of obs.B given in the first part of Table 2 \\

\end{small}
\label{spectra}
\end{table}

\subsection{Phase resolved spectroscopy}

After correcting  the time of arrival of the source events to the
Solar System barycenter, we carried out  a standard timing
analysis based on folding and phase fitting. The source period was
clearly detected in all the observations, with the values given in
Table \ref{obslog}.  Note that the pulse period reported in
Mereghetti et al. (2004) for obs. B is wrong.  As a consequence,
also the corresponding pulsed fraction (computed by fitting a
sinusoid to the background subtracted light curve)  is slightly
larger than that previously reported. The folded light curves of
the three observations are shown in Fig.2 (left panel). They are
characterized by a single broad pulse of nearly sinusoidal shape
and different levels of modulation in the three observations. We
have computed  the pulsed fraction by fitting the background
subtracted pulse profiles   with a   constant plus a sinusoid. The
pulsed fraction of each observation, defined as the ratio between
the semi-amplitude of the sinusoid and the constant, is plotted as
a function of the source flux in the right panel of Fig.~2.



\begin{figure*}[tb]
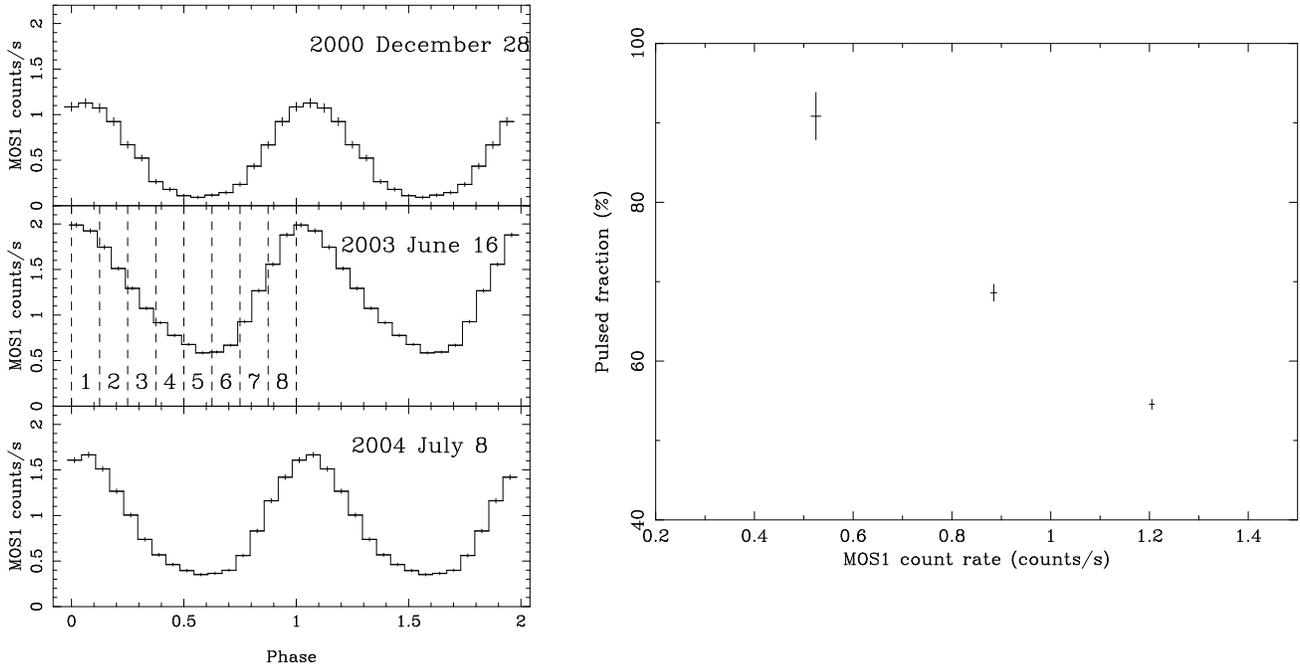

\mbox{}
 \vspace{9.5cm}
  \includegraphics{vefolds.ps}
  \includegraphics{pf.ps}
 \caption[]{{\it Left panel:} Background subtracted pulse  profiles of \ee in
 the 0.6--10 keV energy range obtained with the MOS1 (the only EPIC camera operated in the same mode for all observations).
The phase intervals used for the pulse phase spectroscopy of obs. B are indicated.
{\it Right panel:} Pulsed fraction as function of the average count rate.
 }
 \label{f1}
\end{figure*}

\begin{figure*}[tb]
\mbox{}
\vspace{10.5cm}
\includegraphics{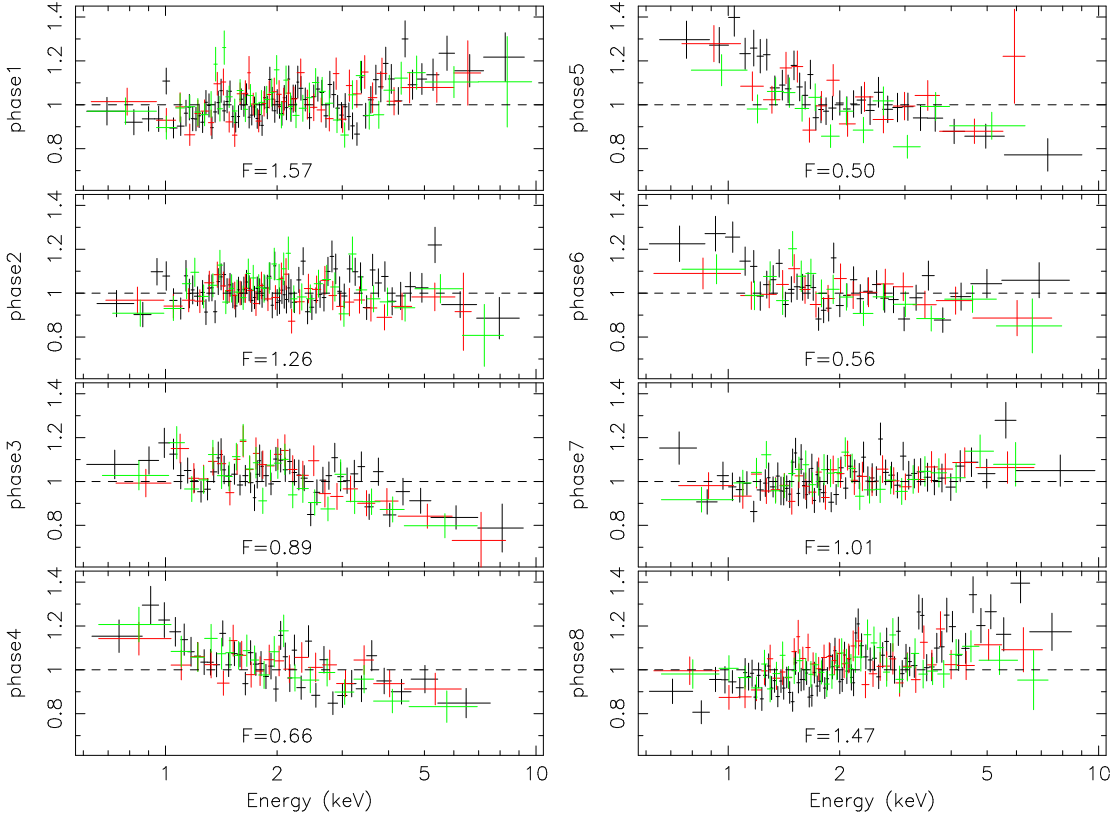}
 \caption[]{Phase resolved spectroscopy of obs.B. The figures show the ratio between
the phase resolved spectra and the best-fit power-law plus blackbody model
rescaled by a factor F (indicated in each panel).
The data (black  PN, red MOS1, green  MOS2) have been graphically rebinned to emphasize the spectral trend.}
 \label{resphas}
\end{figure*}

Phase resolved spectroscopy was performed first by dividing  the
data of obs. B in 8 phase intervals as indicated in Fig. 2. For
each interval a spectrum  was extracted and analyzed as described
above for the phase averaged spectra. Again, to study the phase
dependent variations, we adopted for simplicity the power law plus
blackbody model. The spectra of the different intervals were first
fitted keeping the parameters fixed at  the best-fit values of the
phase averaged spectrum, and allowing only the overall
normalization factor to vary. The corresponding residuals and
normalization factors, shown in Fig.\ref{resphas},  clearly
indicate that the spectrum is softer at pulse minimum and harder
at maximum. To model these spectral variations, as a first step we
allowed the normalizations of the two spectral components to vary
independently. Then, also the blackbody temperature and power law
photon index were left free to vary. The resulting best-fit values
are plotted in Fig. 4. As it was found in the comparison of the
phase averaged spectra, the spectral variations can be modelled
equally well by a change in the blackbody or in the power law
component. What is interesting is that, in any case, a change in
the relative normalization of the two components is required.

The same kind of  analysis applied to the other two
observations\footnote{due to the small number of  counts  we used
only three phase intervals for obs. A} led to similar results.
An example of this, for the case of fixed kT$_{BB}$ (fourth column
of Fig. 4), is shown in Fig. 5.

\begin{figure*}[tb]
\mbox{}
\vspace{9.5cm}
 \includegraphics{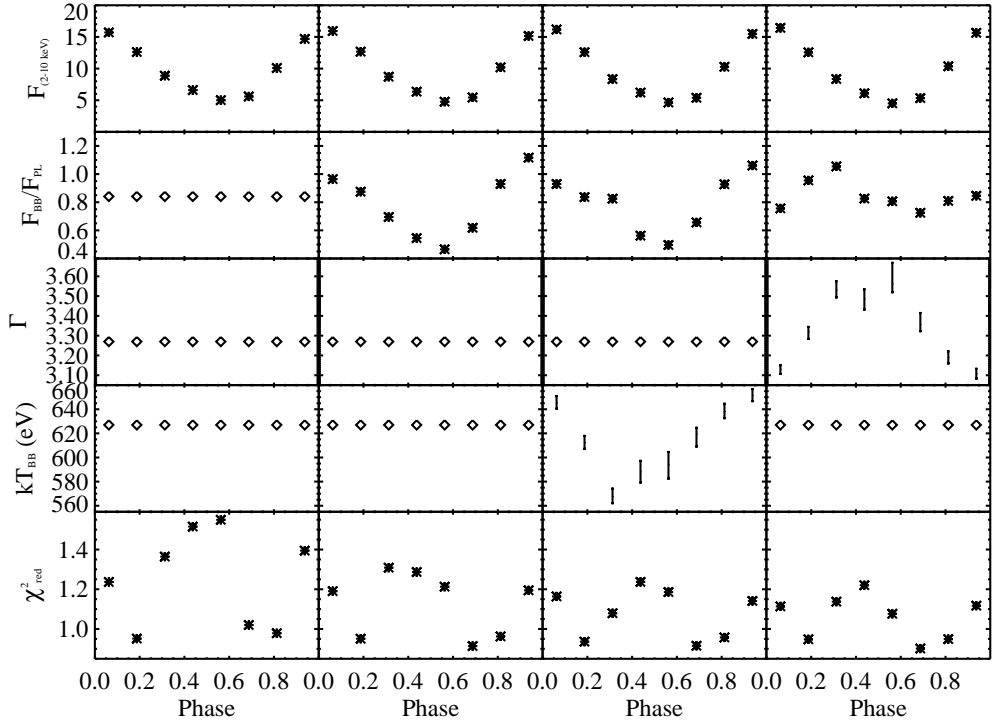}
 \caption[]{Best-fit parameters for the phase resolved spectra of obs. B.
From top to bottom: the observed 2-10 keV flux (in units of
10$^{-12}$ erg cm$^{-2}$ s$^{-1}$), the ratio of the 0.6-10 keV
(absorbed) flux in the blackbody and power law components, the
power law photon index, the blackbody temperature and the reduced
$\chi^2$.
The diamonds indicate that the parameter is fixed to the best-fit value
of the phase averaged spectrum. All the error bars represent 1$\sigma$ errors.}

 \label{pps2003}
\end{figure*}

\begin{figure*}[tb]
\mbox{}
\vspace{9.5cm} \includegraphics{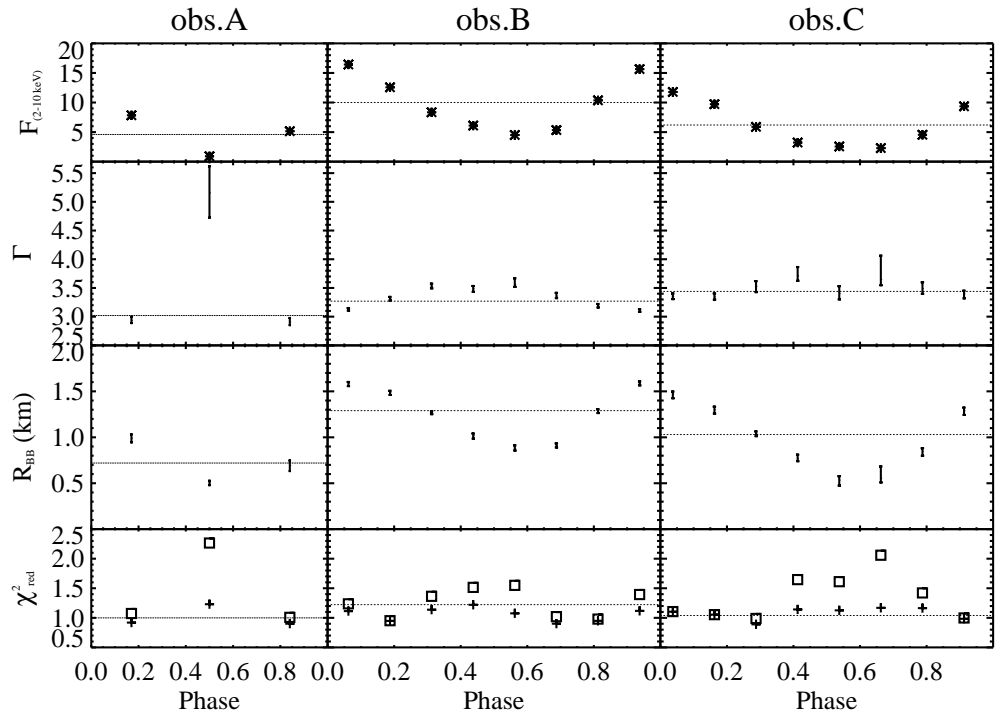} \caption[]{Best-fit parameters for
the phase resolved spectra. The horizontal lines indicate the
values for the phase averaged spectra. F$_{2-10 keV }$ is the PN
observed flux (in units of 10$^{-12}$ erg cm$^{-2}$ s$^{-1}$).
N$_H$ and kT$_{BB}$ have been fixed at the best-fit values of
obs.B.
The error bars are at 1$\sigma$. In the bottom panels the reduced $\chi^2$ values (crosses) are
compared to those  (squares) obtained by fixing
also $\Gamma$ and the
ratio of the normalizations of the power law and
blackbody to  the best-fit values of obs.B.
}
 \label{f1}
\end{figure*}

\begin{figure*}[tb]
\mbox{} \vspace{9.5cm} \includegraphics{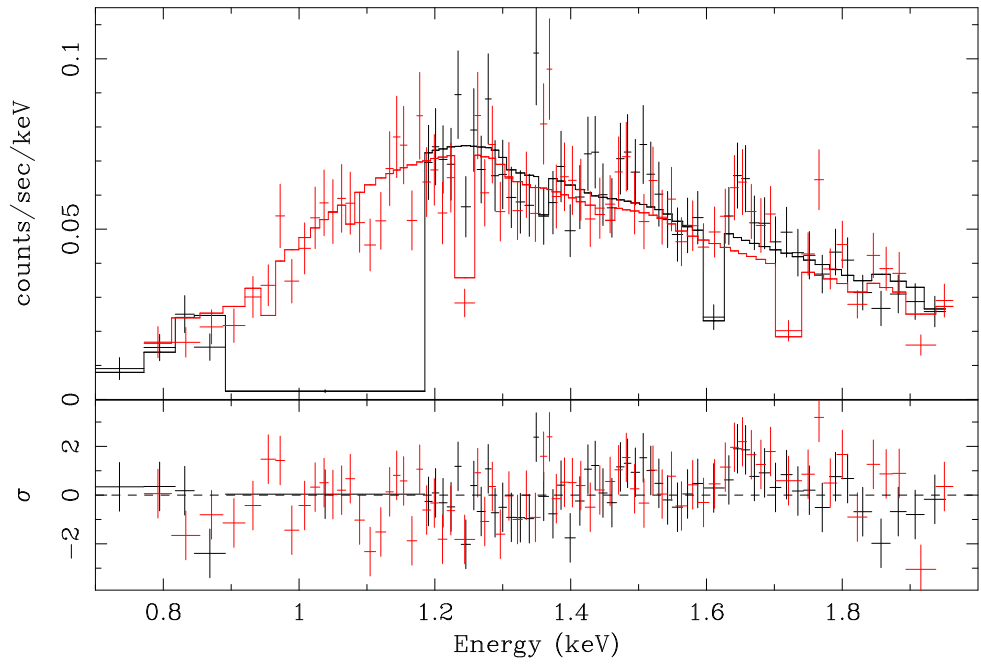}
 \caption[]{First order RGS1 (black) and RGS2 (red) spectra of \ee during obs. B. The sharp drops in the data and model are due to the gaps between chips.
No RGS1 data are present in the $\sim$0.9--1.2 keV range because of the failure of one CCD.
 The lower panel shows the residuals
 (in standard deviation units) with respect to the EPIC best-fit power law plus blackbody
 model.}
 \label{rgs}
\end{figure*}

\subsection{Search for spectral features}

We searched  for spectral features in the EPIC spectra by first
considering lines with a Gaussian profile and width narrower than
the instrumental energy resolution (i.e. fixing $\sigma$=0) and
then extending the search to broader lines ($\sigma$=100 eV). No
spectral features were significantly detected. Since some
deviations from the best fit models, which are most likely to be
attributed to small calibration uncertainties were present in the
residual, we adopted a conservative confidence level of 4$\sigma$
for the upper limits. These are given for the equivalent widths of
either emission or absorption lines in Table 4.

For obs. B, thanks to the long exposure time and high source flux,
also the RGS instrument could be used for the spectral analysis.
The RGS covers only the 0.35--2.5 keV range and has a smaller
effective area than EPIC, but thanks to
its higher energy resolution ($\sim$5 eV at 1 keV compared to 80
eV of EPIC)
it is very efficient in the search for narrow lines in the soft
X--ray range. After a standard extraction of the RGS source and
background spectra and the generation of the corresponding
response files, the first order spectra\footnote{the higher order
spectra do not have enough counts for a meaningful analysis.} of
the RGS1 and RGS2 instruments were rebinned in order to have at
least 50 counts per bin. The spectra were then jointly fitted
using the blackbody plus power law best fit parameters obtained
with EPIC, leaving only a normalization factor for each RGS unit
as free parameter. The values of these factors are 0.99$\pm$0.04
for RGS1 and 0.90$\pm$0.03 for RGS2, which are fully compatible
with the known cross-calibration uncertainties. However, this fit
is not formally acceptable ($\chi_{red}^2=$1.22/127) and some
excess in the residuals is visible around 1.65 keV (see
Fig.\ref{rgs}). In fact, a significantly better fit
($\chi_{red}^2=$1.05/125) is obtained with the addition of a
narrow emission line centered at 1.653$^{+0.007}_{-0.005}$ keV and
with equivalent width 21$\pm$7 eV (90\% confidence level). However
the reality of this line is not confirmed by the higher statistics
EPIC data:
the 4 $\sigma$ upper limit on the equivalent width of a narrow
emission line at 1.653 keV in the EPIC spectrum of obs. B is 6.7
eV.


\begin{table}[htbp]
\begin{center}
  \caption{Upper limits (at 4 $\sigma$) on spectral features in the EPIC
  spectra of \ee from  obs. B.
  To derive these values we used narrow ($\sigma$ = 0 eV) and
  broad ($\sigma$ = 100 eV) Gaussian lines in emission or absorption.}

    \begin{tabular}[c]{ccc}
\hline
Energy range        & $\sigma$ (eV) & Equivalent width \\
\hline
1--2 keV & 0 & $<$ 10 eV \\
& 100 & $<$ 20 eV \\
2--4 keV & 0 & $<$ 20 eV \\
& 100 & $<$ 30 eV \\
4--5 keV & 0& $<$ 30 eV \\
& 100 & $<$ 55 eV \\
5--6 keV & 0 & $<$ 60 eV \\
& 100 & $<$ 60 eV \\
6--7 keV & 0 & $<$ 90 eV \\
& 100 & $<$ 150 eV \\
7--8 keV & 0 & $<$ 150 eV \\
& 100 & $<$ 220 eV \\
\hline

 \end{tabular}
\end{center}
\label{lines}
\end{table}

The search for spectral features was also extended to the EPIC
phase resolved spectra (the smaller number of counts does not
allow to use also the RGS for phase-resolved spectroscopy).
No narrow features were detected in the 8 phase resolved EPIC
spectra described in Section 2.2. The  4$\sigma$ upper limits for
emission or absorption lines are typically a factor $\sim$2--5
higher than those reported in Table 4, depending on the pulse
phase interval and energy range.

\section{Discussion}


The target of opportunity observation of July 2004 (obs. C) was
motivated by the detection of one burst from the direction of \ee\
with the RXTE satellite (Kaspi et al. 2004) and carried out nine
days after this event. No bursts were seen during the \xmm\
observation, nor any other sign indicating a particularly
increased level of activity: the flux of 1E~1048.1--5937, 6.2$\times10^{-12}$
erg cm$^{-2}$ s$^{-1}$, was still $\sim$25\% higher than the
typical values observed before the 2001 outburst (see Fig.2 in
Mereghetti et al. 2004 and Fig.1 in Gavriil \& Kaspi 2004), but
not as high as that measured in obs. B. 
We find statistically significant differences in the spectra of 
the three \xmm\ observations, however such differences are
not related in a monotonic way with the source luminosity: when
the flux was at the highest level (obs. B) the spectral hardness
was intermediate (see Fig.1).

A coherent pattern is instead present in the relation between flux
and pulsed fraction: the latter decreases when the source
brightens, as was first reported by Mereghetti et al. (2004) based
only on two observations.  This  is confirmed by obs. C as shown
in the right panel of Fig.~2, where, to avoid any cross
calibration uncertainty, we directly compare the flux and pulsed
fraction (P$_{F}$) measured with the same detector  in the three
observations.
%
The existence of an empirical anti-correlation between luminosity
and pulsed fraction, independent on its theoretical explanation,
must be taken into account in the interpretation of the RXTE
results for this source. The luminosity and fluence obtained with
this satellite, which can only measure the pulsed component of the
flux, are clearly underestimated if the decrease in pulsed
fraction during the outbursts is not taken into account. We
estimate that  total energy release of   the flares peaking in
November 2001 and June 2002 is at least the double of the value (2
and 20 $\times10^{40}$ ergs respectively)  derived assuming a
constant P$_{F}$=.94 by Gavriil \& Kaspi (2004).


%
Our results for the phase averaged spectroscopy are substantially
in agreement with previous findings for this and other AXPs,
except that the high statistics of obs. B allowed us to reject the
fit with two blackbody components.  We  also explored spectral
fits using a simple Comptonization model. The fit with the
Comptonization model alone is not satisfactory. The addition,
however, of a second, softer blackbody component, resulted in an
acceptable fit, both for the phase-averaged, and for the
phase-resolved spectra. Quite interestingly the estimate of the
radiating area for the cooler component is in this case
compatible 
with the entire star surface when the distance uncertainty is taken
into account. Spectra at different
phases can be fitted keeping this emitting area fixed at the value
derived from the phase-averaged analysis and this results only in
a modest variation of the cool blackbody temperature, $\approx
10\%$, and a modulation of the hot blackbody emitting area. The
smaller area associated with the hotter blackbody emission may be
suggestive of a scenario in which a magnetically active region
produces a hotter region contributing a substantial part of the
luminosity.
The accelerated  high-energy particles, besides heating this
region may upscatter the soft photons giving rise to the
Comptonized spectrum.


The implementation of a detailed Comptonization model to fit the X--ray spectrum
is beyond the scope of our analysis. Therefore, we have used the
approximated model described in Section 2.1, which is based on a constant scattering 
cross-section. 
The scattering cross-section in a 
strongly magnetized medium is anisotropic and
quite different for ordinary (O) and extraordinary (X) photons (e.g. 
Ventura 1979; Gonthier et al. 2000).
In our case the typical incident photon energy is $\approx 0.5$~keV, well below the
resonance at the electron cyclotron energy $\mathrm{E_B}\sim
11.6(\mathrm{B}/10^{12}\, \mathrm{G})\, \mathrm{keV}$. In the
low-frequency limit ($\mathrm{E}\ll\mathrm{E_B}$), the cross section for ordinary
photons is $\sigma_\mathrm{O}\sim\sigma_\mathrm{T}\sin^2\theta$, where
$\sigma_\mathrm{T}$ is the Thomson cross-section. Due to relativistic beaming,
scattering occurs at an incident angle
$\theta\la 1/(2\gamma)$ in the particle frame and therefore 
$\sigma_\mathrm{O}\sim\sigma_\mathrm{T}/(4\gamma^2)$. For relativistic electrons of a given energy,
the cross section for ordinary photons is therefore constant. The cross section for 
extraordinary photons, instead, peaks at $\theta\sim (\mathrm{E/E_B})^{1/2}$, where 
$\sigma_\mathrm{X}\sim\sigma_\mathrm{T}(\mathrm{E/E_B})$. 
It means that, taking $\mathrm{E/E_B}\approx 10^{-3}$, it dominates over 
$\sigma_\mathrm{O}$ for $\gamma\ga 15$. In this case the use of a constant 
cross-section is possible because we have checked numerically that its dependence
on the photon spectrum has a negligible effect on the Comptonized spectrum.

The present model does not allow to univocally determine the properties of the
Comptonizing medium, since the fitted parameter $\alpha$ depends on both the 
scattering depth and the particle energy. 
However, with the presently derived values of the photon 
index, $\alpha\sim 5$, once the Comptonizing particles are assumed to be 
relativistic electrons, the medium is optically thin and the optical depth
is strongly anticorrelated to the particle energy 
($\tau^B_{\mathrm es}\sim \gamma^{-8}$).
For example,  if $\gamma\sim 5$,  the magnetic depth is 
$\tau^B_{\mathrm es}\sim 10^{-6}$, to which corresponds a
non-magnetic Thomson depth $\sim 10^{-4}$, while if $\gamma\sim 50$, 
$\tau^B_{\mathrm es}$ drops to $10^{-14}$, corresponding to a non-magnetic Thomson 
depth of $\sim 10^{-11}$.

Thanks to the high statistics of our observations we could set
strong upper limits on the presence of  spectral lines (in view of
the EPIC results we do not consider the possible feature at 1.65
keV seen in the RGS as really present in the source).
If interpreted in terms of magneto-dipolar
braking, the spin-down rate measured with RXTE at the time of obs.
B (Gavriil \& Kaspi 2004) yields a dipolar field $\approx8\times
10^{14} \, {\rm G}$ and at these field strenghts cyclotron, free-bound and 
free-free features are expected to lie in the observed energy band. 
On the other hand, present estimates of their detectability 
in the presence of superstrong 
magnetic fields still suffer from a number of approximations  (Zane et al.
2002, Ho et al.~2003). Several effects, e.g. fast rotation and
magnetic smearing, have been suggested to explain the absence of
spectral lines in other sources (e.g. Braje \& Romani 2002).
Although the former does not apply to the present source, both
cyclotron and atomic lines are sensitive to the local field
strength and the contributions of  surface elements with different
values of $B$ may under certain conditions wash out spectral
features. Ho \& Lai~(2004) also suggested that at super-strong
field strengths ($B \ga 10^{14}$~G) vacuum polarization may
efficiently suppress spectral lines, but current treatment of this
QED effect is too crude to make a definite statement. 
The absence of lines 
may also suggest either that
the magnetic field at the neutron star surface is larger than the
large scale dipolar component (i.e. that $B \geq 15. \times 10^{14} \,
{\rm G} $) or, alternatively,  that the value inferred by the
spin-down rate is overestimated (i.e. that $B \leq   10^{14}
\, {\rm G} $). As discussed by Thompson et al. (2002), this value
must be regarded as an upper limit: a twisted magnetosphere may
lead to a reduction up to an order of magnitude in the inferred
polar value of the magnetic field, while the estimate of the spin
down age is unaffected.

Phase resolved spectroscopy has now been carried out for most
AXPs. The earlier observations obtained with the ASCA and BeppoSAX
satellites gave some indications for phase-dependent spectral
variations in 4U 0142$+$614 (Israel et al. 1999), 1E~2259$+$586
(Corbet et al. 1995, Parmar et al. 1998), \ee (Oosterbroek et al.
1998), and a more significant evidence for 1RXS J1708--4009
(Israel et al. 2001). These indications were later confirmed by
observations with higher statistics carried out with \xmm\ (Tiengo
et al. 2002, Woods et al. 2004, G\"{o}hler et al. 2004) and {\it
Chandra} (Morii et al. 2003). In most of these studies the
blackbody plus power-law was adopted and the spectral variations
described as changes in one or more of the model parameters. The
results reported here give strong   evidence  that the spectrum of
\ee is harder at pulse maximum. However the interpretation of this
result is not straightforward since the values of the spectral
parameters, e.g. blackbody temperature or photon index, are linked
in a complex way to the overall spectral shape/hardness.
%
%
As  we have shown in the previous section, these variations can be
modelled in different ways. For example (see Fig.\ref{pps2003}) if
the photon index and blackbody  temperature are fixed, the ratio
$F_{BB}/F_{PL}$ is proportional to the total intensity. In turn,
this means that the blackbody contribution at the pulse maximum is
more important than that at the pulse minimum, i.e. that the
blackbody component has a pulsed fraction larger than the power
law. This is what is expected if the thermal component originates
from the star surface and the non-thermal one from a more extended
magnetosphere, and may lead to suggest that the observed spectral
changes are simply due to the partial occultation of an the
emitting regions as the star rotates, with the effect being more
prominent for the surface emission than for the magnetosphere. On
the other hand, better fits are obtained if also the photon index
or the temperature are free to vary (third and fourth columns of
Fig.~4). In this respect, probably the most interesting result is
that we found impossible to reproduce the  phase-dependent
variations keeping one of the two components completely fixed
(i.e. shape and normalization). This fact indicates a coupling
between the two components, or alternatively, it strengthens a
scenario in which the emission below $\sim$10 keV is dominated by
a single physical mechanism with a spectrum more complex than a
blackbody or a power-law.


\section{Summary}

The main conclusions derived from the analysis and comparison of
the three \xmm\ observations of \ee\ reported here are the
following:

\begin{itemize}

\item  long term flux variations of a factor \gtsima2 are
accompanied by only minor, though statistically significant,
changes in the source spectral properties

\item the   pulsed fraction is anti-correlated with the source
luminosity, decreasing from $\sim$90\% to $\sim$55\% in
correspondence of a twofold increase in flux. This should be
properly taken into account when the source energetics are
inferred by measurements of the pulsed flux

\item spectral variations are clearly present as a function of the
pulse phase, with  the hardest spectrum at pulse maximum; their
interpretation in terms of simple changes in two component models
is not unique

\item there is no evidence for  absorption and/or emission lines
in the spectrum with equivalent width larger than 10-220 eV,
depending on energy and width

\item the spectrum is consistent with a two component model
consisting of a  blackbody emission from an emitting area
compatible with the whole neutron star surface and a hotter
Comptonized blackbody, possibly coming from a relatively large
region of the star surface heated by non-thermal magnetospheric
particles also responsible for the scattering.

\end{itemize}


\end{document}